\documentclass[a4paper,12pt]{article}
\usepackage[left=20mm,right=20mm,top=20mm,bottom=30mm]{geometry}
\usepackage{microtype}
\usepackage{amsmath}
\usepackage{amssymb}
\usepackage{mathtools}
\usepackage{hyperref}
\usepackage{setspace}
\usepackage{tikz}
\usepackage{cite}
\usepackage{multirow}
\usepackage[affil-it]{authblk}
\usepackage{caption}

\usetikzlibrary{calc}
\usetikzlibrary{math}
\usetikzlibrary{arrows.meta}

\hypersetup{
  colorlinks,
  linkcolor={blue!50!black},
  citecolor={blue!50!black},
  urlcolor={blue!50!black}
}

\newcommand{\abs}[1]{\lvert #1 \rvert}

\DeclareMathOperator{\arcsinh}{arcsinh}

\begin{document}

\title{Production of heavy charged particles in proton-proton ultraperipheral
collisions at the Large Hadron Collider: survival factor}

\author[1]{S.~I.~Godunov}
\author[1]{V.~A.~Novikov}
\author[2,3]{A.~N.~Rozanov}
\author[1]{M.~I.~Vysotsky}
\author[1]{E.~V.~Zhemchugov~\thanks{evgenii.zhemchugov@cern.ch}}

\affil[1]{\small I.E. Tamm Department of Theoretical Physics, Lebedev Physical
Institute, Moscow 119991 Russia}
\affil[2]{\small Institute for Theoretical and Experimental Physics,
Moscow 117218 Russia}
\affil[3]{\small Centre de Physique des Particules de Marseille, CPPM,
Aix-Marseille Universite, CNRS/IN2P3, Marseille F-13288 France}

\date{}

\maketitle

\begin{abstract}
Ultraperipheral collisions of high energy protons are a source of approximately
real photons colliding with each other.  Photon fusion can result in production
of yet unknown charged particles in very clean events. The cleanliness of such
an event is due to the requirement that the protons survive during the
collision.  Finite sizes of the protons reduce the probability of such outcome
compared to point-like particles. We calculate the survival factors and cross
sections for the production of heavy charged particles at the Large Hadron
Collider.
\end{abstract}

\section{Introduction}

Protons accelerated at the LHC are the sources of strong electromagnetic fields
carried by almost real equivalent photons. The idea to consider the LHC as a
photon-photon collider was analyzed in detail long ago~\cite{prep163-299,
hep-ph-0112211, hep-ph-0112239, hep-ex-0201034, hep-ph-0304301, nucl-ex-0502005,
hep-ph-0611042, 0706.3356, 0810.1400, 1104.0571, 1404.0896, 1601.07001,
1607.05095, 1610.06647, 1708.09836}. If new charged particles exist, they should
be produced in ultraperipheral collisions (UPC) of protons. The corresponding
cross section is model independent and is determined by masses and electric
charges of these particles.

Finite sizes of protons modify the equivalent photons approximation developed
for point-like colliding particles. First, the equivalent photons flux is
reduced by the elastic form factor. Second, the protons should miss each other
to avoid inelastic interaction and loss of coherence. The corresponding
reduction of the total cross section of the UPC production of definite final
state depends on a single variable $r_p \sqrt{s} / \gamma$, where $\sqrt{s}$ is
the invariant mass of the created system, $r_p$ is the proton radius, and
$\gamma$ is the Lorentz factor of each
proton~\cite{prd42-3690}.\footnote{Dimensionless reduction of the total cross
section depends on two dimensionful quantities: $r_p$ and $\sqrt{s}$. Photon
energy $\omega$ in EPA is always divided by $\gamma$. The corresponding scaling
behaviour follows. When additional scales are introduced (e.g. cuts), this
simple behaviour is no longer correct.} This is the reason why the survival
factor, being at the level of few percents in the case of the production of a
muon pair, considerably diminishes the cross section for the production of heavy
particles.

To take into account the reduction of the photon flux, we use the
Weizs\"acker-Williams equivalent photons approximation (EPA) in the impact
parameter space. Our approach is close to that of Ref.~\cite{prd42-3690}, where
one can find references to the earlier literature devoted to this problem.
However, we are taking into account elastic form factors explicitely, while in
Ref.~\cite{prd42-3690} the nucleus interior is completely excluded: integrations
over impact parameters $b_i$ go from $r_p$ to infinity.

In the framework of electroweak theory, diagrams in which one or both
intermediate photons are replaced by $Z$~bosons also contribute to the particles
production. At the amplitude level, replacement of a photon by $Z$ results in
the suppression factor of the order of $(\Lambda_\text{QCD} / m_Z)^2 \sim
10^{-6}$ as soon as we demand that the proton remains intact after emitting the
$Z$~boson. Consequently, contributions of intermediate $Z$ into the UPC
processes considered in the present paper are completely negligible.

In Section~\ref{s:epa} explicit formulae which take into account finite sizes of
the colliding particles are derived. They are applied in
Section~\ref{s:chargino} to the production of heavy charged particles and in
Section~\ref{s:atlas} to the production of $\mu^+ \mu^-$ pairs in proton-proton
collisions. We conclude in Section~\ref{s:conclusion}.

\section{EPA and the survival factor}

\label{s:epa}

The spectrum of equivalent photons of a proton with the Lorentz factor $\gamma$
is~\cite{prc47-2308, prc48-2011, 1806.07238}
\begin{equation}
  n(b, \omega)
  = \frac{\alpha}{\pi^2 \omega}
    \left[
      \int\limits_0^\infty
        \frac{F(q_\perp^2 + \omega^2 / \gamma^2)}
             {q_\perp^2 + \omega^2 / \gamma^2}
        \, J_1(b q_\perp)
        \, q_\perp^2
        \, \mathrm{d} q_\perp
    \right]^2,
  \label{epa-spectrum/ip}
\end{equation}
where $n(b, \omega) \, \mathrm{d}^2 b \, \mathrm{d} \omega$ is the number of
photons with the energy $\omega$ moving through a point of space with the impact
parameter $\vec b$, $\alpha$ is the fine structure constant, $q_\perp$ is the
transverse photon momentum, $q_\perp^2 + \omega^2 / \gamma^2 \equiv Q^2$ is the
photon virtuality, $F(Q^2)$ is the Dirac form factor of the proton, $J_1(z)$ is
the Bessel function of the first kind.

With the help of the relation
\begin{equation}
  \int\limits_0^\infty J_1(ax) \, J_1(bx) \, x \, \mathrm{d} x
  = \frac{1}{a} \, \delta(a - b),
\end{equation}
integration of $n(b, \omega)$ over the whole impact parameter space
$\mathrm{d}^2 b$ results in the following well known formula for the spectrum of
equivalent photons:
\begin{equation}
  n(\omega)
  = \int n(b, \omega) \mathrm{d}^2 b
  = \frac{2 \alpha}{\pi \omega}
    \int\limits_0^\infty
      \left[
        \frac{F(q_\perp^2 + \omega^2 / \gamma^2)}
             {q_\perp^2 + \omega^2 / \gamma^2}
      \right]^2
      q_\perp^3
    \, \mathrm{d} q_\perp.
  \label{epa-spectrum}
\end{equation}

The proton Dirac form factor is~\cite{pr119-1105, prep550-1}
\begin{equation}
  F(Q^2) = \frac{G_E(Q^2) + \tau G_M(Q^2)}{1 + \tau},
\end{equation}
where $G_E(Q^2)$ and $G_M(Q^2)$ are known as the Sachs electric and magnetic
form factors, $\tau = Q^2 / 4 m_p^2$, $m_p$ is the proton mass. In
ultraperipheral collisions $Q^2 \lesssim \Lambda^2_\text{QCD} \sim
0.04~\text{GeV}^2$~\cite{1806.07238}, so $\tau \lesssim 0.01 \ll 1$, and the
magnetic contribution can usually be neglected. However, our calculations have
shown that the magnetic form factor contribution to the photon-photon luminosity
is at the same order of magnitude as the survival factor contribution, so we
keep it in the following.

One common description for the electric and magnetic form factors is the dipole
approximation
\begin{equation}
  G_E(Q^2) = \frac{1}{(1 + Q^2 / \Lambda^2)^2},
  \quad
  G_M(Q^2) = \frac{\mu_p}{(1 + Q^2 / \Lambda^2)^2},
\end{equation}
where $\mu_p = 2.7928473508(85)$ is the proton magnetic
moment~\cite{1507.07956}, and the standard value of $\Lambda^2$ is
$\Lambda_\text{std}^2 = 0.71~\text{GeV}^2$. Relatively recent experimental data
disfavor the dipole approximation~\cite{1307.6227}. However, according to
Fig.~10(a) of Ref.~\cite{1307.6227}, the discrepancy between the fit to the
experimental data and the standard dipole approximation is at the level of a few
percent for $Q^2 \lesssim 0.04~\text{GeV}^2$, with the fit being below the
approximation. The slope of $G_E(Q^2)$ at $Q^2 = 0$ is related to the charge
radius of proton:
\begin{equation}
  r_p^2 = -6 \, \frac{\mathrm{d} G_E(Q^2)}{\mathrm{d} Q^2}.
\end{equation}
With the CODATA value of $r_p = 0.8751(61)$~fm~\cite{1507.07956} (which is in
agreement with the value obtained in Ref.~\cite{1307.6227}, see Eq.~(52)), we
get another estimation of $\Lambda^2$:
\begin{equation}
  \Lambda_\text{CODATA}^2 = 12 / r_p^2 = 0.61~\text{GeV}^2.
\end{equation}
We use this value in the following calculations.\footnote{\label{ft:a1} Though
dipole form factor with $\Lambda_\text{CODATA}$ gives much better description of
experimental data than with $\Lambda_\text{std}$ for $Q^2 \lesssim
0.04~\text{GeV}^2$, let us note that in the region $Q^2\gtrsim
0.04~\text{GeV}^2$ data points are above dipole approximation with
$\Lambda_\text{CODATA}$. This may lead to the actual cross section being greater
by $\sim 2$--3\%.}

\begin{figure}[!bt]
  \includegraphics{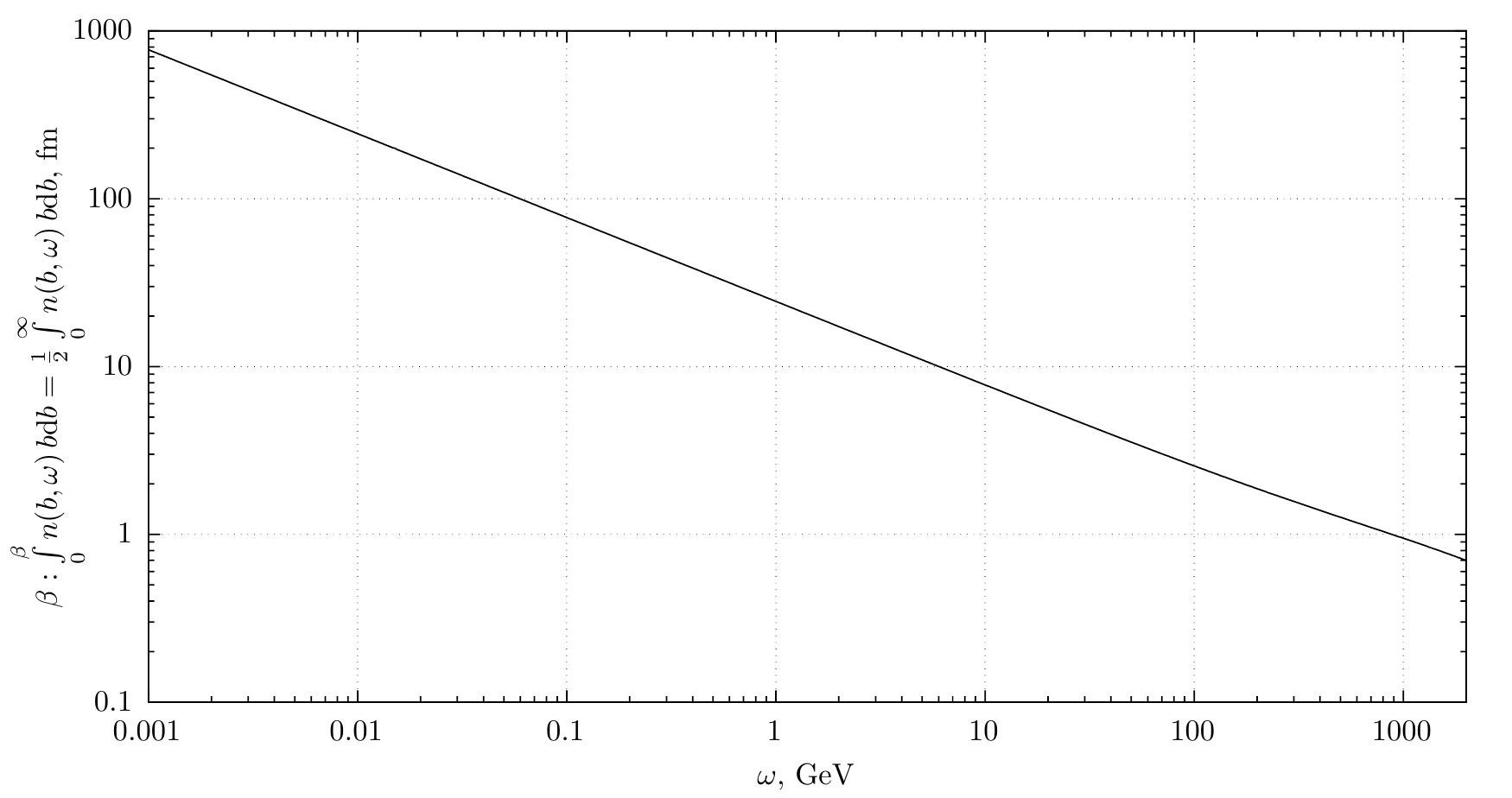}
  \caption{Median of the equivalent photon space distribution with respect to
  photon energy for a $6.5$~TeV proton. For a given photon energy, there is the
  same amount of photons in the region $r < \beta$ as is in the region $r >
  \beta$, where $r$ is the distance to the proton center.}
  \label{f:median}
\end{figure}

Application of the dipole approximation to Eq.~\eqref{epa-spectrum/ip} results
in
\begin{multline}
  \begin{split}
    n(b, \omega)
    &= \frac{\alpha}{\pi^2 \omega}
       \left[
           \frac{\omega}{\gamma} K_1 \left( \frac{b \omega}{\gamma} \right)
         - \left(
               1
             + \frac{(\mu_p - 1) \frac{\Lambda^4}{16 m_p^4}}{
                 \left( 1 - \frac{\Lambda^2}{4 m_p^2} \right)^2
               }
           \right)
           \sqrt{\Lambda^2 + \frac{\omega^2}{\gamma^2}}
           \, K_1 \left( b \sqrt{\Lambda^2 + \frac{\omega^2}{\gamma^2}} \right)
    \right. \\ &\qquad \left.
         + \frac{(\mu_p - 1) \frac{\Lambda^4}{16 m_p^4}}{
             \left( 1 - \frac{\Lambda^2}{4 m_p^2}  \right)^2
           }
           \sqrt{4 m_p^2 + \frac{\omega^2}{\gamma^2}}
           \, K_1 \left( b \sqrt{4 m_p^2 + \frac{\omega^2}{\gamma^2}} \right)
    \right. \\ &\qquad \left.
         - \frac{1 - \frac{\mu_p \Lambda^2}{4 m_p^2}}
                {1 - \frac{\Lambda^2}{4 m_p^2}}
         \cdot
           \frac{b \Lambda^2}{2}
           \, K_0 \left( b \sqrt{\Lambda^2 + \frac{\omega^2}{\gamma^2}} \right)
       \right]^2,
  \end{split}
\end{multline}
where $K_0(z)$, $K_1(z)$ are the modified Bessel functions of the second kind
(the Macdonald functions). Integration over $\mathrm{d}^2 b$ yields
\begin{equation}
  \begin{split}
    n(\omega)
    &= \frac{\alpha}{\pi \omega}
       \left\{
           \left( 1 + 4 u - 2 (\mu_p - 1) \frac{u}{v} \right)
           \ln \left( 1 + \frac{1}{u} \right)
    \right. \\ &\qquad \left. {}
         + \frac{\mu_p - 1}{(v - 1)^4} \left[
               \frac{\mu_p - 1}{v - 1} (1 + 4 u + 3 v)
             - 2 \left( 1 + \frac{u}{v} \right)
           \right]
           \ln \frac{u + v}{u + 1}
         - \frac{24 u^2 + 42 u + 17}{6 (u + 1)^2}
    \right. \\ &\qquad \left. {}
         + (\mu_p - 1) \,
           \frac{
             6 u^2 (v^2 - 3 v + 3) + 3 u (3 v^2 - 9 v + 10) + 2 v^2 - 7 v + 11
           }{
             3 (u + 1)^2 (v - 1)^3
           }
    \right. \\ &\qquad \left. {}
         - (\mu_p - 1)^2 \,
           \frac{
             24 u^2 + 6 u (v + 7) - v^2 + 8 v + 17
           }{
             6 (u + 1)^2 (v - 1)^4
           }
       \right\},
  \end{split}
\end{equation}
where
\begin{equation}
  u = \left( \frac{\omega}{\Lambda \gamma} \right)^2, \ 
  v = \left( \frac{2 m_p}{\Lambda} \right)^2.
\end{equation}
To get a grasp on the photon distribution $n(b, \omega)$, see
Fig.~\ref{f:median}, where the median of the equivalent photon space
distribution is depicted. The median is defined as such distance $\beta$ that
\begin{equation}
  \int\limits_{b < \beta} \mathrm{d}^2 b \, n(b, \omega)
  = \int\limits_{b > \beta} \mathrm{d}^2 b \, n(b, \omega).
\end{equation}

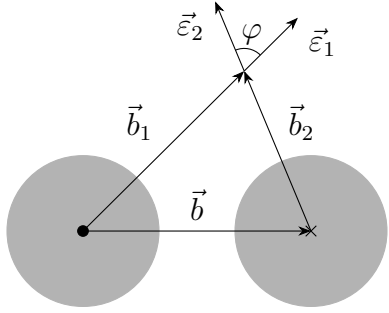
\begin{figure}
  \centering
  \begin{tikzpicture}[>=Stealth]
    \tikzmath{\R = 1; \l = 0.07;}

    \coordinate (Cl) at (-1.5 * \R, 0);
    \coordinate (Cr) at ( 1.5 * \R, 0);

    \filldraw [black!30!white] (Cl) circle [radius=\R];
    \filldraw (Cl) circle [radius=\l];
    \filldraw [black!30!white] (Cr) circle [radius=\R];
    \draw     ($(Cr) + (\l, \l)$) -- ($(Cr) - (\l, \l)$);
    \draw     ($(Cr) + (\l, -\l)$) -- ($(Cr) + (-\l, \l)$);

    \draw [->] (Cl) -- node [above] {$\vec b$} (Cr);

    \coordinate (A) at ($(Cl) + (45:3 * \R)$);

    \draw [->] (Cl) -- node [above left]  {$\vec b_1$} (A);
    \draw [->] (Cr) -- node [above right] {$\vec b_2$} (A);

    \draw [->] (A) -- +(45:1)    node [below right] {$\vec \varepsilon_1$};
    \draw [->] (A) -- +(112.5:1) node [below left]  {$\vec \varepsilon_2$};
    \draw      ($(A) + (45:0.3)$) arc (45:112.5:0.3);
    \node at ($(A) + (78.75:0.5)$) {$\varphi$};
  \end{tikzpicture}
  \caption{
    An ultraperipheral collision of two protons moving perpendicular to the
    figure plane. $\vec b_1$ and $\vec b_2$ are impact parameters of a point in
    space at which photons collide relative to the corresponding protons. $b =
    \abs{\vec b_1 - \vec b_2}$ is the impact parameter of the collision.
    $\vec \varepsilon_1$ and $\vec \varepsilon_2$ are the photon polarization
    vectors.
  }
  \label{f:upc}
\end{figure}

Consider production of a system $X$ in an ultraperipheral collision of two
protons. The collision geometry is presented in Fig.~\ref{f:upc}.
The cross section is
\begin{equation}
  \sigma(pp \to pp X)
  = \int\limits_0^\infty \mathrm{d} \omega_1
    \int\limits_0^\infty \mathrm{d} \omega_2
    \int\limits_{b_1 > \hat r} \mathrm{d}^2 b_1
    \int\limits_{b_2 > \hat r} \mathrm{d}^2 b_2
    \, \sigma(\gamma \gamma \to X)
    \, n(b_1, \omega_1)
    \, n(b_2, \omega_2)
    \, P(b),
  \label{pp->ppX-ip}
\end{equation}
where $\sigma(\gamma \gamma \to X)$ is the $X$ production cross section in
photon fusion, and $P(b)$ is the probability for the protons to survive in a
collision with the impact parameter $b = \abs{\vec b_1 - \vec b_2}$.  The
integration domain over $\vec b_1$ and $\vec b_2$ is limited by the
conditions $b_1 > \hat r$, $b_2 > \hat r$.  The value of $\hat r$ depends on the
nature of the produced system $X$. If $X$ contains strongly interacting
particles, as the first approximation one can let $\hat r = r_p$, the proton
charge radius; more accurate treatment of this case requires modification of
Eq.~\eqref{pp->ppX-ip} to account for the probability for the particles produced
inside a proton to escape. In the following we consider production of particles
which do not participate in strong interactions, and we neglect rescattering
processes through electromagnetic or weak interactions, so we set $\hat r = 0$.

Photons in ultraperipheral collisions are polarized in the directions shown in
Fig.~\ref{f:upc}. Let $\sigma_\parallel(\gamma \gamma \to X)$ and
$\sigma_\perp(\gamma \gamma \to X)$ be the cross sections for the $\gamma \gamma
\to X$ reaction with the photons polarization vectors parallel or perpendicular.
These cross sections depend only on the product $\omega_1 \omega_2$, therefore
they can be factored out through the change in the integration variables from
$\omega_1$, $\omega_2$ to $s = 4 \omega_1 \omega_2$, $y = \tfrac12 \ln
\tfrac{\omega_1}{\omega_2}$ ($\sqrt{s}$ is the invariant mass of $X$, $y$ is the
rapidity of $X$)~\cite{npa518-786}:
\begin{equation}
  \sigma(pp \to pp X)
  = \int\limits_0^\infty \mathrm{d} s
    \left[
     \, \sigma_\parallel(\gamma \gamma \to X)
     \, \frac{\mathrm{d} L_\parallel}{\mathrm{d} s}
     +  \sigma_\perp(\gamma \gamma \to X)
     \, \frac{\mathrm{d} L_\perp}{\mathrm{d} s}
    \right],
  \label{xsection}
\end{equation}
where
\begin{equation}
  \begin{aligned}
    \frac{\mathrm{d} L_\parallel}{\mathrm{d} s}
    &= \frac14
       \int\limits_{-\infty}^\infty \mathrm{d} y
       \int\limits_{b_1 > \hat r} \mathrm{d}^2 b_1
       \int\limits_{b_2 > \hat r} \mathrm{d}^2 b_2
       \, n \left( b_1, \tfrac{\sqrt{s}}{2} \, \mathrm{e}^y \right)
       \, n \left( b_2, \tfrac{\sqrt{s}}{2} \, \mathrm{e}^{-y} \right)
       \, P(b)
       \, \cos^2 \varphi,
    \\
    \frac{\mathrm{d} L_\perp}{\mathrm{d} s}
    &= \frac14
       \int\limits_{-\infty}^\infty \mathrm{d} y
       \int\limits_{b_1 > \hat r} \mathrm{d}^2 b_1
       \int\limits_{b_2 > \hat r} \mathrm{d}^2 b_2
       \, n \left( b_1, \tfrac{\sqrt{s}}{2} \, \mathrm{e}^y \right)
       \, n \left( b_2, \tfrac{\sqrt{s}}{2} \, \mathrm{e}^{-y} \right)
       \, P(b)
       \, \sin^2 \varphi
  \end{aligned}
  \label{luminosity/ip}
\end{equation}
are the photon-photon luminosities for the corresponding relative photon
polarizations in UPC of two protons, $\varphi$ is the angle between the vectors
$\vec b_1$, $\vec b_2$ as is shown in Fig.~\ref{f:upc}. In the case when the
cross section does not depend on the photons polarization, Eq.~\eqref{xsection}
simplifies to
\begin{equation}
  \sigma(pp \to pp X)
  = \int\limits_0^\infty \mathrm{d} s
    \, \sigma(\gamma \gamma \to X)
    \, \frac{\mathrm{d} L}{\mathrm{d} s},
  \label{xsection-unpolarized}
\end{equation}
where $L = L_\parallel + L_\perp$. 

When the non-electromagnetic interactions between the protons are neglected,
$P(b) = 1$, and the luminosities can be expressed through the integrated
spectrum~\eqref{epa-spectrum}:
\begin{equation}
    \left. \frac{\mathrm{d} L_\parallel}{\mathrm{d} s} \right\rvert_{P=1}
  = \left. \frac{\mathrm{d} L_\perp}{\mathrm{d} s} \right\rvert_{P=1}
  = \frac12 \left. \frac{\mathrm{d} L}{\mathrm{d} s} \right\rvert_{P=1}
  = \frac18
    \int\limits_{-\infty}^\infty \mathrm{d} y
       \, n \left( \tfrac{\sqrt{s}}{2} \, \mathrm{e}^y \right)
       \, n \left( \tfrac{\sqrt{s}}{2} \, \mathrm{e}^{-y} \right).
  \label{luminosity}
\end{equation}
Note that in this case the luminosities for different polarizations are equal,
so one can use the simpler formula~\eqref{xsection-unpolarized} with
$\sigma(\gamma \gamma \to X) = \tfrac12 (\sigma_\parallel(\gamma \gamma \to X) +
\sigma_\perp(\gamma \gamma \to X))$ to calculate the cross section. In other
words, when the non-electromagnetic interactions between the colliding particles
are neglected, so can be the polarization of photons.

Ref.\cite{hep-ph-0608271} suggests the following expression for $P(b)$:
\begin{equation}
  P(b) = \left( 1 - \mathrm{e}^{-\frac{b^2}{2 B}} \right)^2.
  \label{upc-probability}
\end{equation}
The parameter $B$ depends on the collision energy $E$. An empirical formula is
provided in Ref.~\cite{1112.3243}:
\begin{equation}
  B = B_0 + 2 B_1 \ln(E / E_0) + 4 B_2 \ln^2(E / E_0)
\end{equation}
with $B_0 = 12~\text{GeV}^{-2}$, $B_1 = -0.22 \pm 0.17~\text{GeV}^{-2}$, $B_2 =
0.037 \pm 0.006~\text{GeV}^{-2}$, $E_0 = 1$~GeV. The validity of this formula was
tested for $E = 7$ and 8~TeV~\cite{1408.5778, 1607.06605} and was found to be in
a good agreement with the experiment. For $E = 13$~TeV we will use $B =
21.1~\text{GeV}^{-2}$.

Substituting~\eqref{upc-probability} into~\eqref{luminosity/ip} and integrating
over the angles, we get
\begin{equation}
  \begin{gathered}
    \begin{multlined}
      \frac{\mathrm{d} L_\parallel}{\mathrm{d} s}
      = \frac{\pi^2}{2}
        \int\limits_{\hat r}^\infty b_1 \, \mathrm{d} b_1
        \int\limits_{\hat r}^\infty b_2 \, \mathrm{d} b_2
        \int\limits_{-\infty}^\infty \mathrm{d} y
        \, n \left( b_1, \tfrac{\sqrt{s}}{2} \, \mathrm{e}^y \right)
        \, n \left( b_2, \tfrac{\sqrt{s}}{2} \, \mathrm{e}^{-y} \right)
        \\  \times
        \left\{
            1
          - 2 \mathrm{e}^{-\frac{b_1^2 + b_2^2}{2 B}}
            \left[
                I_0 \left( \frac{b_1 b_2}{B} \right)
              + I_2 \left( \frac{b_1 b_2}{B} \right)
            \right]
          + \mathrm{e}^{-\frac{b_1^2 + b_2^2}{B}}
            \left[
                I_0 \left( \frac{2 b_1 b_2}{B} \right)
              + I_2 \left( \frac{2 b_1 b_2}{B} \right)
            \right]
        \right\},
    \end{multlined}
    \\
    \begin{multlined}
      \frac{\mathrm{d} L_\perp}{\mathrm{d} s}
      = \frac{\pi^2}{2}
        \int\limits_{\hat r}^\infty b_1 \, \mathrm{d} b_1
        \int\limits_{\hat r}^\infty b_2 \, \mathrm{d} b_2
        \int\limits_{-\infty}^\infty \mathrm{d} y
        \, n \left( b_1, \tfrac{\sqrt{s}}{2} \, \mathrm{e}^y \right)
        \, n \left( b_2, \tfrac{\sqrt{s}}{2} \, \mathrm{e}^{-y} \right)
        \\  \times
        \left\{
            1
          - 2 \mathrm{e}^{-\frac{b_1^2 + b_2^2}{2 B}}
            \left[
                I_0 \left( \frac{b_1 b_2}{B} \right)
              - I_2 \left( \frac{b_1 b_2}{B} \right)
            \right]
          + \mathrm{e}^{-\frac{b_1^2 + b_2^2}{B}}
            \left[
                I_0 \left( \frac{2 b_1 b_2}{B} \right)
              - I_2 \left( \frac{2 b_1 b_2}{B} \right)
            \right]
        \right\},
    \end{multlined}
  \end{gathered}
\end{equation}
where $I_0(z)$, $I_2(z)$ are the modified Bessel functions of the first kind, and the equality
\begin{equation}
  \int\limits_0^\pi
    \mathrm{e}^{z \cos \varphi} \cos(n \varphi)
  \, \mathrm{d} \varphi
  = \pi I_n(z)
\end{equation}
is used.

Survival factor describes the diminishing of the UPC cross section due to
disintegration of the colliding particles. At the amplitude level, it is defined
as the ratio of the cross sections with and without the survival effects
included~\cite{hep-ph-0010163,hep-ph-0201301,1405.0018,1508.02718,1710.11505,2007.12704}:
\begin{equation}
  \left< S_{\gamma \gamma} \right>
  = \frac{
      \int\limits_0^\infty \mathrm{d} \omega_1
      \int\limits_0^\infty \mathrm{d} \omega_2
      \int \mathrm{d}^2 b_1
      \int \mathrm{d}^2 b_2
      \, \sigma(\gamma \gamma \to X)
      \, n(b_1, \omega_1)
      \, n(b_2, \omega_2)
      \, P(b)
    }{
      \int\limits_0^\infty \mathrm{d} \omega_1
      \int\limits_0^\infty \mathrm{d} \omega_2
      \, \sigma(\gamma \gamma \to X)
      \, n(\omega_1)
      \, n(\omega_2)
    }.
\end{equation}
When the dependence of the $\gamma \gamma \to X$ cross section
on the relative polarization of the photons is neglected, one gets the following
expression for the survival factor:
\footnote{In the literature, the survival factor is often denoted as
$\left<S_{\gamma \gamma}\right>^2$, $S_{\gamma \gamma}^2$.}
\begin{equation}
  S_{\gamma \gamma}
  = \frac{\mathrm{d} L / \mathrm{d} s \, \mathrm{d} y}
         {\mathrm{d} L / \mathrm{d} s \, \mathrm{d} y \rvert_{P=1}}
  = \frac{\mathrm{d} L / \mathrm{d} \omega_1 \, \mathrm{d} \omega_2}
         {\mathrm{d} L / \mathrm{d} \omega_1 \, \mathrm{d} \omega_2 \rvert_{P=1}}.
  \label{survival-gg}
\end{equation}
$S_{\gamma \gamma}$ depends on the invariant mass $\sqrt{s}$ and rapidity $y$ of
the system produced or, equivalently, on the photons energies $\omega_1$,
$\omega_2$. In the following we will use the definition of the survival factor
with the luminosities integrated with respect to rapidity:
\begin{equation}
  S
  = \frac{\mathrm{d} L / \mathrm{d} s}
    {\mathrm{d} L / \mathrm{d} s \rvert_{P=1}}.
  \label{survival}
\end{equation}
In the case of the production of a pair of charged fermions with the mass $m$,
Eqs.~\eqref{survival-gg}, \eqref{survival} are valid for $s \gg m^2$, see
Eq.~\eqref{breit-wheeler} in the next section.

\section{Production of heavy charged particles}

\label{s:chargino}

\begin{figure}[!p]
  \centering
  \includegraphics{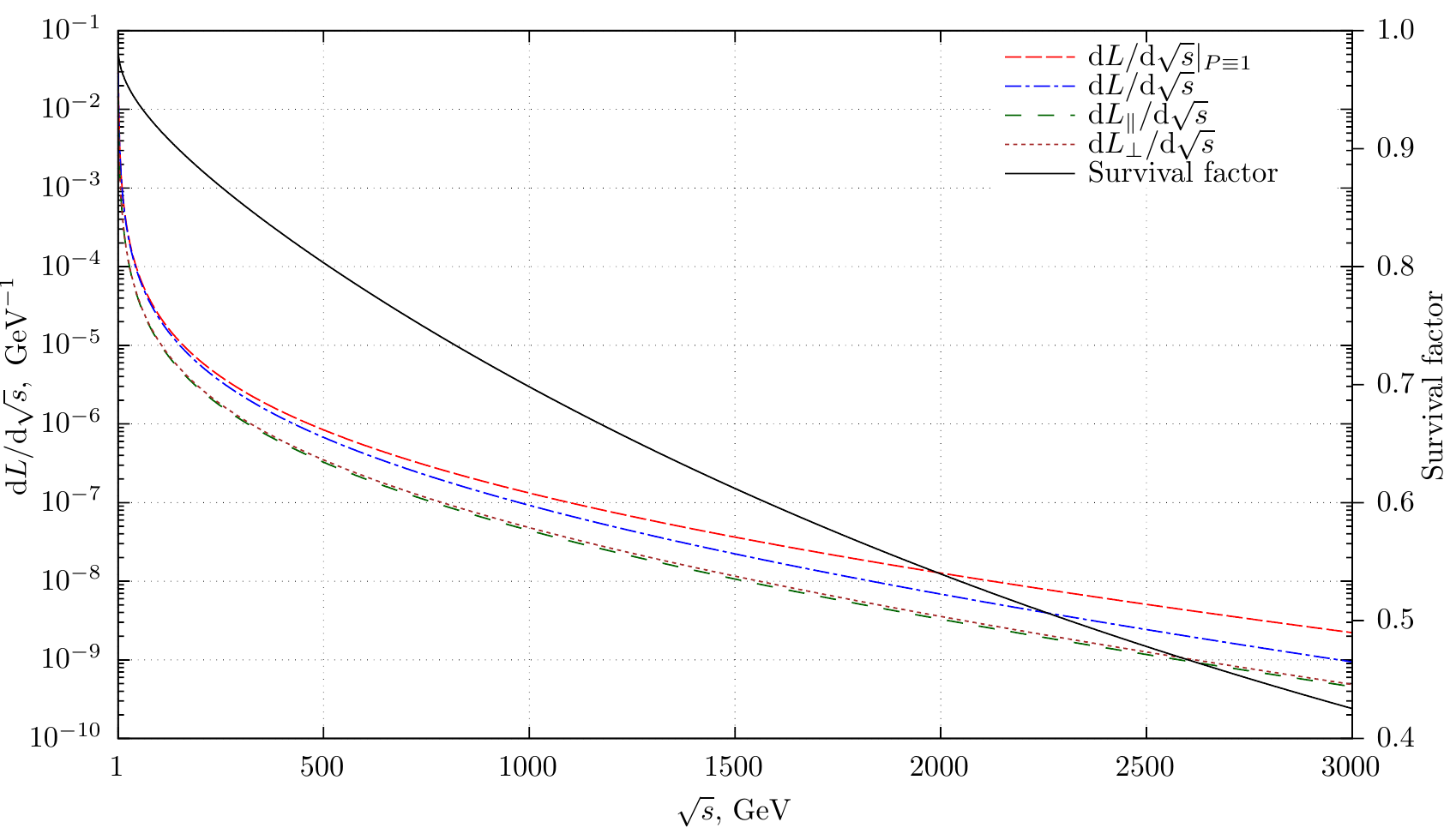}
  \caption{
    Survival factor $S$~\eqref{survival} (right axis) and corresponding
    equivalent photon-photon luminosities (left axis) in $pp$ collisions with
    the energy 13~TeV. Here $\sqrt{s}$ is the invariant mass of the produced
    system.
  }
  \label{f:survival}
\end{figure}

Results of calculations of photon-photon luminosities in proton-proton
collisions with the energy 13~TeV through
Eqs.~\eqref{luminosity/ip},~\eqref{luminosity} are presented in
Fig.~\ref{f:survival} along with the survival factor~\eqref{survival}.\footnote{
 Ref.~\cite{npa518-786} argues that at large invariant masses $\mathrm{d}
 L_\parallel / \mathrm{d} s > \mathrm{d} L_\perp / \mathrm{d} s$. In
 proton-proton collisions with the energy 13~TeV this transition occurs at
 $\sqrt{s} \sim 7$~TeV.
}
The difference between the luminosities of photons with parallel and
perpendicular polarizations is below 10\%.

\begin{figure}[!p]
  \centering
  \includegraphics{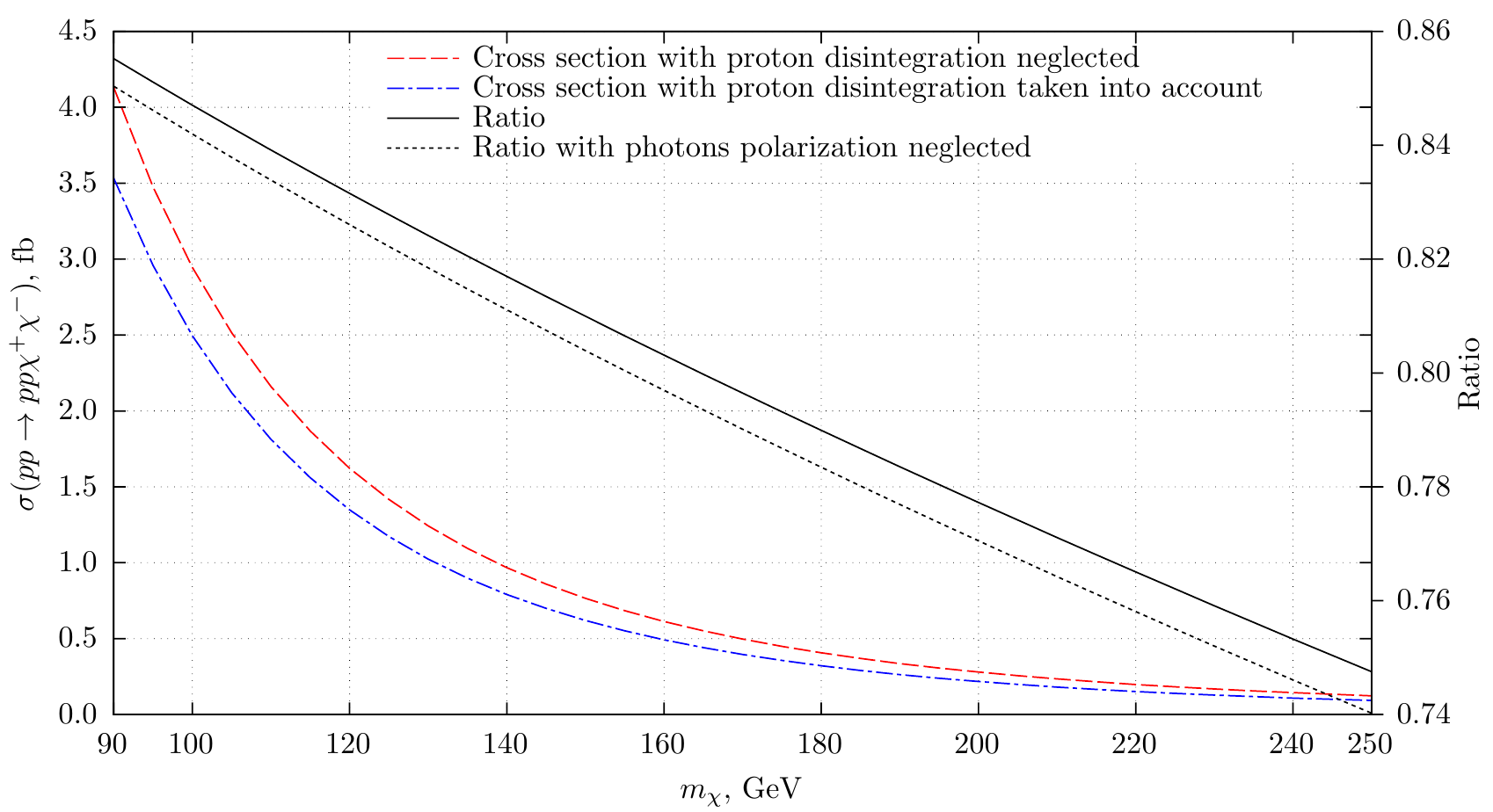}
  \caption{
    Cross sections (left axis) and their ratio (right axis) for the production of
    a pair of charged particles $\chi^+ \chi^-$ in $pp$ collisions with the
    energy 13~TeV with respect to the particle mass. The black dotted line
    (right axis) is the ratio of the cross sections when the polarization of
    photons in UPC is neglected. The corresponding cross section is not shown
    because in this plot it is barely distinguishable from the dash-dotted line.
  }
  \label{f:chargino}
\end{figure}

Production of a pair of charged fermions $\chi^+ \chi^-$ with the charge 1 and
the mass $m_\chi$ is described by the Breit-Wheeler cross
sections~\cite{pr46.1087}:
\begin{equation}
  \begin{aligned}
    \sigma_\parallel(\gamma \gamma \to \chi^+ \chi^-)
    &= \frac{4 \pi \alpha^2}{s}
       \left[
         \left( 1 + \frac{4 m_\chi^2}{s} - \frac{12 m_\chi^4}{s^2} \right)
         \ln \frac{1 + \sqrt{1 - 4 m_\chi^2 / s}}
                  {1 - \sqrt{1 - 4 m_\chi^2 / s}}
         - \left( 1 + \frac{6 m_\chi^2}{s} \right)
           \sqrt{1 - \frac{4 m_\chi^2}{s}}
       \right],
    \\
    \sigma_\perp(\gamma \gamma \to \chi^+ \chi^-)
    &= \frac{4 \pi \alpha^2}{s}
       \left[
         \left( 1 + \frac{4 m_\chi^2}{s} - \frac{4 m_\chi^4}{s^2} \right)
         \ln \frac{1 + \sqrt{1 - 4 m_\chi^2 / s}}
                  {1 - \sqrt{1 - 4 m_\chi^2 / s}}
         - \left( 1 + \frac{2 m_\chi^2}{s} \right)
           \sqrt{1 - \frac{4 m_\chi^2}{s}}
       \right].
  \end{aligned}
  \label{breit-wheeler}
\end{equation}
These formulae are substituted into Eq.~\eqref{xsection} for $X = \chi^+ \chi^-$.
The resulting cross section is shown in Fig.~\ref{f:chargino}.

In Ref.~\cite{1906.08568} we propose a model-independent approach to the search
of charged heavy long-lived particles produced in ultraperipheral collisions at
the LHC.  The corresponding cross section is presented in Fig.~2 of
Ref.~\cite{1906.08568}.\footnote{
  In our calculations in Ref.~\cite{1906.08568} we used $\Lambda^2 =
  \Lambda^2_\text{std} = 0.71~\text{GeV}^2$ and neglected the magnetic component
  of the proton form factor.
} From Fig.~\ref{f:chargino} we see that the cross section obtained in
Ref.~\cite{1906.08568} decreases due to proton disintegration by approximately
15\% for $m_\chi = 100$~GeV. For heavier $\chi$ the decrease is larger according
to the scaling of the survival factor by the variable $r_p \sqrt{s} / \gamma$
(heavier $\chi$ corresponds to larger $\sqrt{s}$ which has the same effect as
making proton radius larger for fixed $m_\chi$).

Experimental data are often constrained by cuts on the phase space imposed to
take into account detector geometry and to improve the signal to background
ratio. Common requirements are for the particles produced to have transverse
momentum $p_T > \hat p_T$ and pseudorapidity $\abs{\eta} < \hat \eta$, where
$\hat p_T$, $\hat \eta$ are the cut values, $\eta = -\ln \tan(\theta / 2)$,
$\theta$ is the angle between the momentum of the produced particle and the
proton beam. The differential fiducial cross section for the production of a
$\chi^+ \chi^-$ pair is then
\begin{multline}
  \frac{\mathrm{d} \sigma_\text{fid.}(pp \to pp \chi^+ \chi^-)}{\mathrm{d} s}
  \\
  = \int\limits_{
      \max \left(
        \hat p_T,
        \frac{\sqrt{s}}{2 \cosh \hat \eta}
        \sqrt{1 - \frac{4 m_\chi^2}{s}}
      \right)
    }^{\frac{\sqrt{s}}{2} \sqrt{1 - \frac{4 m_\chi^2}{s}}}
      \mathrm{d} p_T
      \left[
        \frac{\mathrm{d} \sigma_\parallel(\gamma \gamma \to \chi^+ \chi^-)}
             {\mathrm{d} p_T}
        \frac{\mathrm{d} \hat L_\parallel}{\mathrm{d} s}
        + 
        \frac{\mathrm{d} \sigma_\perp(\gamma \gamma \to \chi^+ \chi^-)}
             {\mathrm{d} p_T}
        \frac{\mathrm{d} \hat L_\perp}{\mathrm{d} s}
      \right],
  \label{fiducial-xsection}
\end{multline}
where the differential with respect to $p_T$ cross sections are
\begin{equation}
  \begin{aligned}
    \frac{\mathrm{d} \sigma_\parallel(\gamma \gamma \to \chi^+ \chi^-)}{\mathrm{d} p_T}
  & = \frac{8 \pi \alpha^2 p_T}{s (p_T^2 + m_\chi^2)}
      \cdot
      \frac{1 - \dfrac{2 (p_T^4 + 2 m_\chi^4)}{s (p_T^2 + m_\chi^2)}}
           {\sqrt{1 - \dfrac{4 (p_T^2 + m_\chi^2)}{s}}},
  \\
    \frac{\mathrm{d} \sigma_\perp(\gamma \gamma \to \chi^+ \chi^-)}{\mathrm{d} p_T}
  & = \frac{8 \pi \alpha^2 p_T}{s (p_T^2 + m_\chi^2)}
      \cdot
      \frac{1 - \dfrac{2 p_T^4}{s (p_T^2 + m_\chi^2)}}
           {\sqrt{1 - \dfrac{4 (p_T^2 + m_\chi^2)}{s}}},
  \end{aligned}
  \label{xsection/pt-polarized}
\end{equation}
and
\begin{equation}
  \begin{aligned}
    \frac{\mathrm{d} \hat L_\parallel}{\mathrm{d} s}
  & = \frac14
      \iint \mathrm{d}^2 b_1 \, \mathrm{d}^2 b_2
      \, P(b)
      \int\limits_{-\hat y}^{\hat y} \mathrm{d} y
      \, n \left( b_1, \tfrac{\sqrt{s}}{2} \, \mathrm{e}^y \right)
      \, n \left( b_2, \tfrac{\sqrt{s}}{2} \, \mathrm{e}^{-y} \right)
      \, \cos^2 \varphi,
  \\
    \frac{\mathrm{d} \hat L_\perp}{\mathrm{d} s}
  & = \frac14
      \iint \mathrm{d}^2 b_1 \, \mathrm{d}^2 b_2
      \, P(b)
      \int\limits_{-\hat y}^{\hat y} \mathrm{d} y
      \, n \left( b_1, \tfrac{\sqrt{s}}{2} \, \mathrm{e}^y \right)
      \, n \left( b_2, \tfrac{\sqrt{s}}{2} \, \mathrm{e}^{-y} \right)
      \, \sin^2 \varphi
  \end{aligned}
  \label{fiducial-luminosity}
\end{equation}
are the photon-photon luminosities~\eqref{luminosity/ip} that take into account
the cut on the pseudorapidity of the produced particles~\cite{1906.08568}:
\begin{equation}
    \hat y
    = \arcsinh
      \left[
        \frac{\sqrt{s} \, p_T}{2(p_T^2 + m_\chi^2)}
         \left(
           \sinh \hat \eta
           - \sqrt{\cosh^2 \hat \eta + \frac{m_\chi^2}{p_T^2}}
             \cdot
             \sqrt{1 - \frac{4 (p_T^2 + m_\chi^2)}{s}}
         \right)
      \right].
  \label{y-bound}
\end{equation}

\begin{figure}[!tb]
  \centering
  \includegraphics{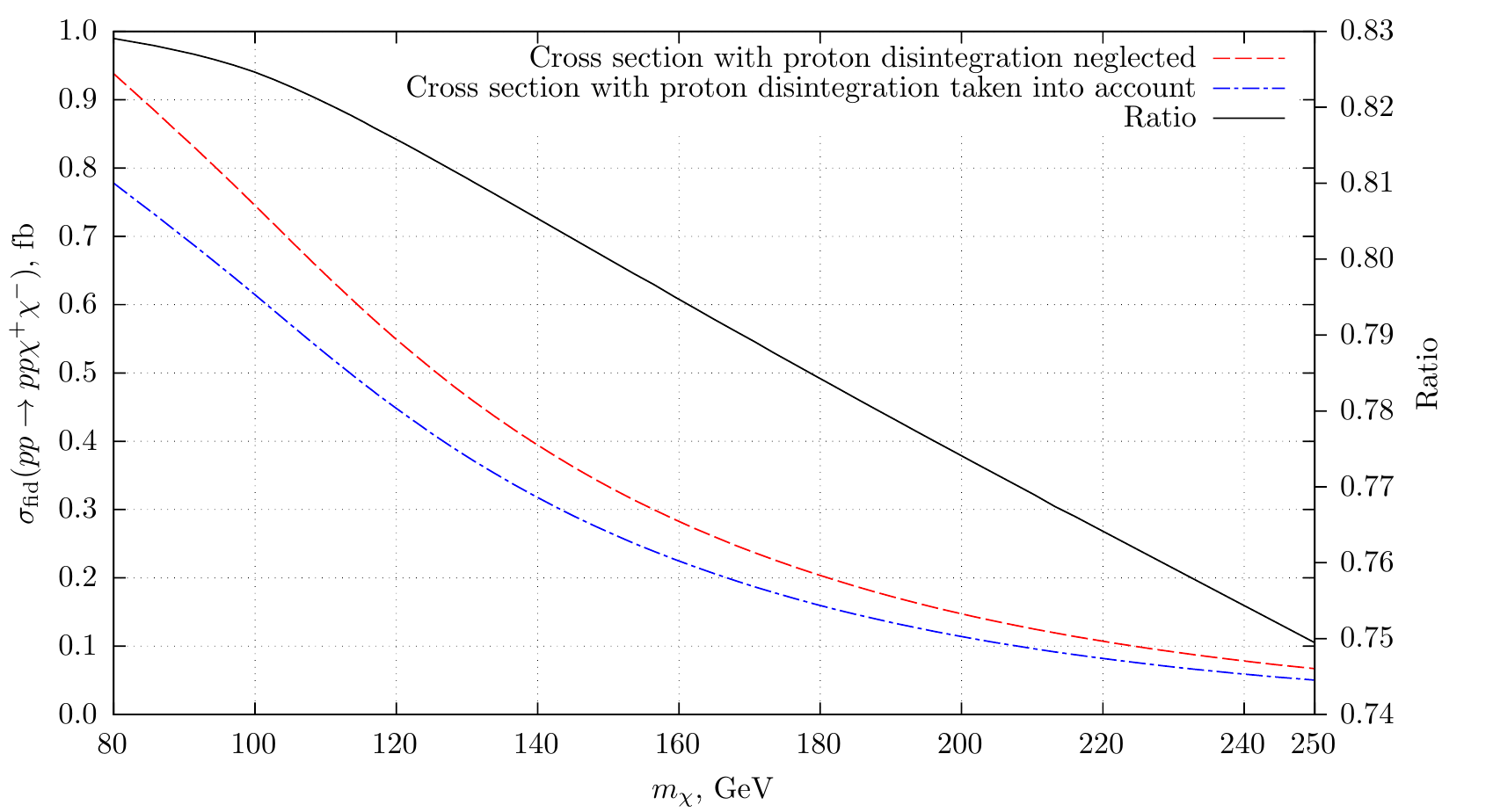}
  \caption{
    Cross sections (left axis) and their ratio (right axis) for the production
    of a pair of charged particles $\chi^+ \chi^-$ in $pp$ collisions with the
    energy 13~TeV with respect to the particle mass. Both particles are required
    to have transverse momentum greater than 20~GeV and pseudorapidity less than
    $2.5$; both protons are required to hit the forward detectors.
  }
  \label{f:fiducial}
\end{figure}

Another common constraint specific to ultraperipheral collisions is the
requirement for the protons to hit the forward detectors. Forward detectors are
special purpose detectors located at $\approx 200$~m down the beam from both
sides of the central detectors of the ATLAS and CMS
collaborations~\cite{atlas-tdr-024-2015, totem-tdr-003}. To hit a forward
detector, a proton has to lose $1.5$--$15$\% of its energy which in the case of
13~TeV collisions corresponds to photon energies from $\omega_\text{min} =
97.5$~GeV to $\omega_\text{max} = 975$~GeV. To apply this constraint to the
calculation of the differential fiducial cross
section~\eqref{fiducial-xsection}, integration limits in
Eq.~\eqref{fiducial-luminosity} have to be modified:
\begin{equation}
  \begin{aligned}
    \frac{\mathrm{d} \hat L_\parallel}{\mathrm{d} s}
  & = \frac14
      \iint \mathrm{d}^2 b_1 \, \mathrm{d}^2 b_2
      \, P(b)
      \int\limits_{
        \max\left(
          -\hat y,
          \frac12 \ln \frac{4 \omega_\text{min}^2}{s},
          \frac12 \ln \frac{s}{4 \omega_\text{max}^2}
        \right)
      }^{
        \min\left(
          \hat y,
          \frac12 \ln \frac{4 \omega_\text{max}^2}{s},
          \frac12 \ln \frac{s}{4 \omega_\text{min}^2}
        \right)
      } \mathrm{d} y
      \, n \left( b_1, \tfrac{\sqrt{s}}{2} \, \mathrm{e}^y \right)
      \, n \left( b_2, \tfrac{\sqrt{s}}{2} \, \mathrm{e}^{-y} \right)
      \, \cos^2 \varphi,
  \\
    \frac{\mathrm{d} \hat L_\perp}{\mathrm{d} s}
  & = \frac14
      \iint \mathrm{d}^2 b_1 \, \mathrm{d}^2 b_2
      \, P(b)
      \int\limits_{
        \max\left(
          -\hat y,
          \frac12 \ln \frac{4 \omega_\text{min}^2}{s},
          \frac12 \ln \frac{s}{4 \omega_\text{max}^2}
        \right)
      }^{
        \min\left(
          \hat y,
          \frac12 \ln \frac{4 \omega_\text{max}^2}{s},
          \frac12 \ln \frac{s}{4 \omega_\text{min}^2}
        \right)
      } \mathrm{d} y
      \, n \left( b_1, \tfrac{\sqrt{s}}{2} \, \mathrm{e}^y \right)
      \, n \left( b_2, \tfrac{\sqrt{s}}{2} \, \mathrm{e}^{-y} \right)
      \, \sin^2 \varphi.
  \end{aligned}
\end{equation}
Integration of the differential fiducial cross section should be performed under
the constraint $4 \omega_\text{min}^2 \le s \le 4 \omega_\text{max}^2$. Cross
section for the production of a pair of charged particles for $\hat p_T = 20$~GeV,
$\hat \eta = 2.5$ and under the requirement that both protons hit the forward
detectors is presented in Fig.~\ref{f:fiducial}. For the muons,
$\sigma_\text{fid}(pp \to pp \mu^+ \mu^-) \approx 1.4$~fb ($1.7$~fb) when the
proton disintegration is taken into account (neglected) which gives the ratio
$0.84$. To be detected in a forward detector, a proton should emit a photon with
the energy at least $97.5$ GeV. That is why the ratio does not reach 1 with the
diminishing of the produced particles mass. It can be clearly seen in
Fig.~\ref{f:fiducial} as solid black line approaches constant in the low mass
region.

\section{Comparison to the experimental data}

\label{s:atlas}

In Ref.~\cite{1806.07238}, M.V. and E.Z. have provided an analytical description
in terms of EPA of muon pair production in proton-proton UPC at the LHC
neglecting the proton disintegration. They have compared their results to the
experimental data obtained in Ref.~\cite{1708.04053}. Let us demonstrate the
effect of the survival factor on the description of the results of this
experiment.

The experimental data are constrained with the cuts on the phase space $p_T >
\hat p_T$, $\abs{\eta} < \hat \eta$, where $p_T$ and $\eta$ are the transverse
momentum and pseudorapidity of each muon. In the limit $p_T \gg m_\mu$, the
differential cross sections $\mathrm{d} \sigma_{\parallel, \perp}(\gamma \gamma
\to \mu^+ \mu^-) / \mathrm{d} p_T$ for parallel and perpendicular photon linear
polarizations~\eqref{xsection/pt-polarized} coincide, therefore the differential
fiducial cross section~\eqref{fiducial-xsection} is
\begin{equation}
  \frac{\mathrm{d} \sigma_\text{fid} (pp \to pp \mu^+ \mu^-)}{\mathrm{d} s}
  = \int\limits_{
      \max \left( \hat p_T, \frac{\sqrt{s}}{2 \cosh \hat \eta} \right)
    }^{\sqrt{s} / 2}
    \mathrm{d} p_T
    \, \frac{\mathrm{d} \sigma(\gamma \gamma \to \mu^+ \mu^-)}{\mathrm{d} p_T}
    \, \frac{\mathrm{d} \hat L}{\mathrm{d} s},
 \label{fiducial-xsection-muons}
\end{equation}
where $\hat L = \hat L_\parallel + \hat L_\perp$, and
\begin{equation}
  \frac{\mathrm{d} \sigma(\gamma \gamma \to \mu^+ \mu^-)}{\mathrm{d} p_T}
  = \frac{8 \pi \alpha^2}{s p_T}
    \frac{1 - 2 p_T^2 / s}{\sqrt{1 - 4 p_T^2 / s}}.
\end{equation}
Eq.~\eqref{y-bound} simplifies to
\begin{equation}
  \hat y
  = \hat \eta
  - \frac12 \ln \frac{1 + \sqrt{1 - 4 p_T^2 / s}}{1 - \sqrt{1 - 4 p_T^2 / s}}.
\end{equation}

\begin{figure}[!tb]
  \centering
  \includegraphics{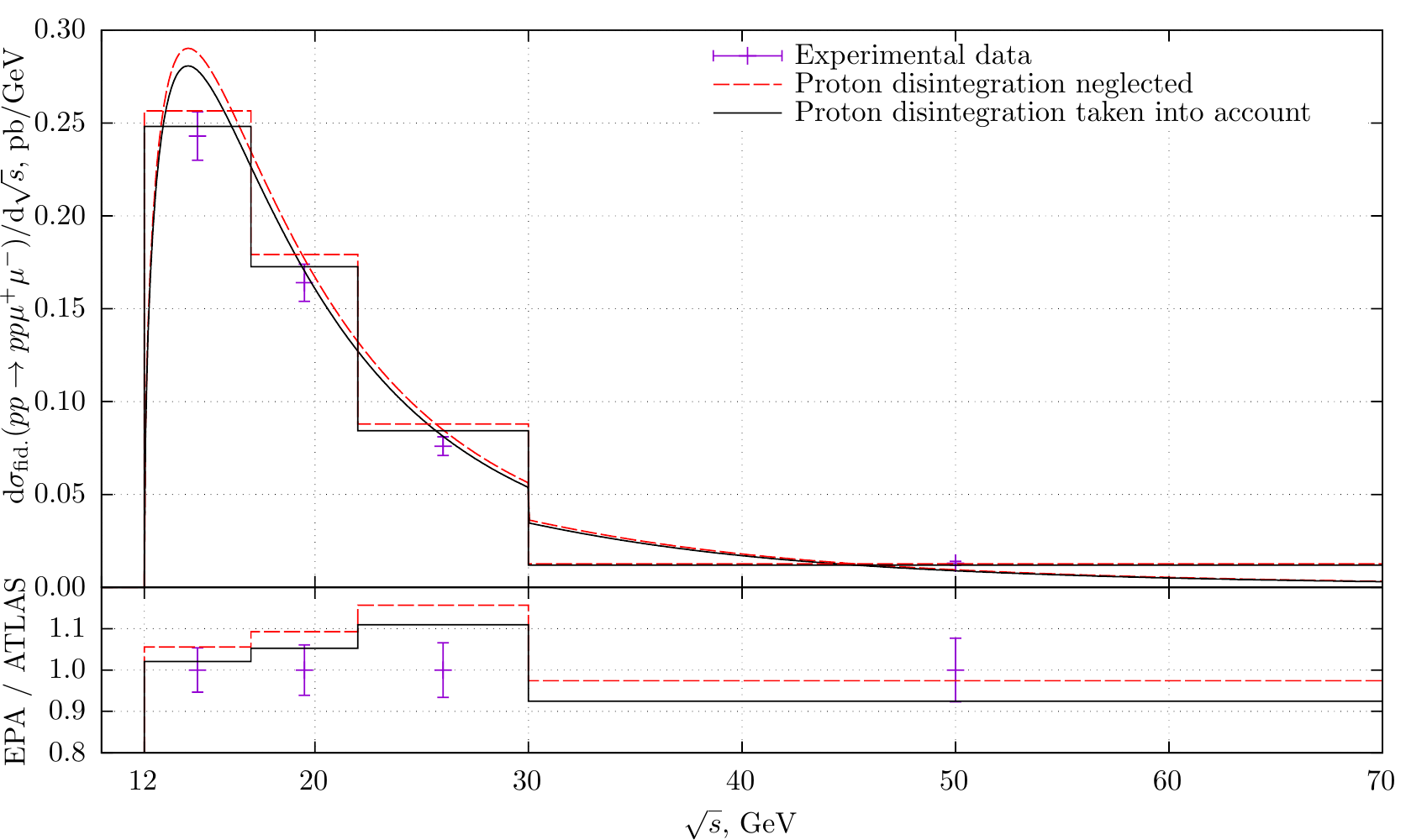}
  \caption{
    {\it Upper plot:} Fiducial cross section of the $pp \to pp \mu^+ \mu^-$
    reaction with the collision energy 13~TeV and the cuts described in
    Table~\ref{t:cuts}. The points are the experimental data from
    Ref.~\cite{1708.04053}. The red dashed line is the differential cross
    section with the proton disintegration probability neglected. The black
    solid line is the differential cross section with the survival factor taken
    into account. The histograms are the cross sections integrated over the
    experimental bins.  {\it Lower plot:} ratio of the calculated cross section
    to the experimental data.
  }
  \label{f:atlas}
\end{figure}

The experimental cuts used in Ref.~\cite{1708.04053} are described in
Table~\ref{t:cuts}.
\begin{table}[!tb]
  \centering
  \caption{Experimental cuts for the fiducial cross section of the $pp (\gamma
  \gamma) \to pp \mu^+ \mu^-$ reaction measured in Ref.~\cite{1708.04053}.}
  \begin{tabular}{ccc}
    Muon pair invariant mass range
    & Muon transverse momentum
    & Muon pseudorapidity
    \\ \hline
    $12~\text{GeV} < \sqrt{s} < 30~\text{GeV}$
    & $p_T > 6$~GeV
    & \multirow{2}{*}{$\abs{\eta} < 2.4$}
    \\
    $30~\text{GeV} < \sqrt{s} < 70~\text{GeV}$
    & $p_T > 10$~GeV
    &
  \end{tabular}
  \label{t:cuts}
\end{table}
The fiducial cross section calculated with the probability for the protons to
survive during the collision~\eqref{upc-probability} is plotted in
Fig.~\ref{f:atlas}.\footnote{
  Note that in our calculations in Ref.~\cite{1806.07238} we used $\Lambda^2 =
  \Lambda_\text{std}^2 = 0.71~\text{GeV}^2$ and neglected the magnetic component
  of the proton form factor. All three corrections: the magnetic form factor
  contribution, the change of $\Lambda^2$ from $\Lambda^2_\text{std}$ to
  $\Lambda^2_\text{CODATA}$, and the survival factor contribution have
  approximately the same magnitude in this experiment, with the former being
  positive and the other two negative.
}

The integrated fiducial cross section in the interval $12~\text{GeV} < \sqrt{s}
< 70~\text{GeV}$ is
\begin{itemize}
  \item experimental value: $3.12 \pm 0.07~\text{(stat.)} \pm
  0.10$~(syst.)~pb~\cite{1708.04053};
  \item with proton disintegration neglected: $3.39$~pb.
  \item with proton disintegration taken into account:
  $3.26$~pb.\footnote{Calculation with the proton form factor provided
  numerically by the A1 collaboration~\cite{1307.6227} (see supplementary
  materials) may result in the cross section $\approx 2$\% greater, see
  footnote~\ref{ft:a1} on page~\pageref{ft:a1}.}
\end{itemize}
Ref.~\cite{1708.04053} also presents EPA predictions made with Monte Carlo
simulations:
\begin{itemize}
  \item {\sc Herwig}~\cite{0803.0883, 1512.01178}: $3.56 \pm 0.05$~pb;
  \item {\sc Herwig} with corrections from Ref.~\cite{1410.2983}: $3.06 \pm 0.05$~pb;
  \item {\sc SuperChic2}~\cite{1508.02718}: $3.45 \pm 0.05$~pb.
\end{itemize}
Our result for the survival factor for the integrated cross section, $\left<
S_{\gamma \gamma} \right> = 3.26~\text{pb} / 3.39~\text{pb} = 0.96$ is the same as in Table~2 of
Ref.~\cite{2104.13392}: $3.43~\text{pb} / 3.58~\text{pb} = 0.96$.

\section{Conclusion}

\label{s:conclusion}

\begin{figure}[!t]
  \centering
  \begin{minipage}{\textwidth}
    \includegraphics{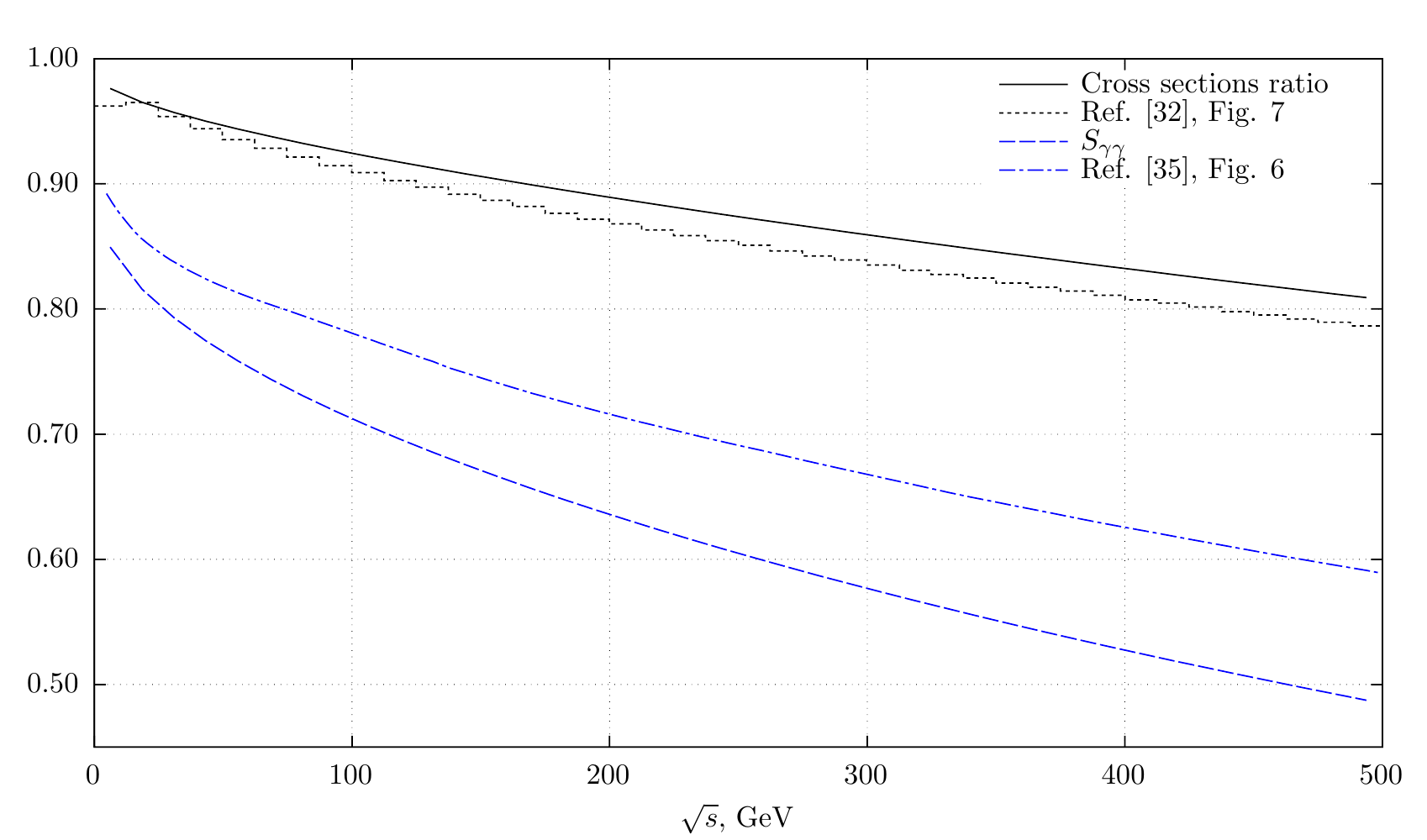}
    \caption{
      Comparison to the results obtained in Refs.\cite{1508.02718, 1410.2983}. The
      upper solid line is the ratio $\frac{\mathrm{d} \sigma_\text{fid.}(pp \to pp
      \, \mu^+ \mu^-) / \mathrm{d} \sqrt{s}}{\mathrm{d} \sigma_\text{fid.}(pp \to
      pp \, \mu^+ \mu^-) / \mathrm{d} \sqrt{s} \rvert_{P=1}}$, where
      $\sigma_\text{fid.}(pp \to pp \, \mu^+ \mu^-)$ is the fiducial cross
      section~\eqref{fiducial-xsection-muons} with the parameters $\hat p_T =
      2.5$~GeV, $\hat \eta = 2.5$ for the $pp$ collision energy 14~TeV; $\sqrt{s}$
      is the invariant mass of the muon pair. It is compared to the histogram
      representing the same value plotted in Ref.~\cite{1508.02718}. The lower
      dashed line is the survival factor $S_{\gamma \gamma}$~\eqref{survival-gg}
      for the collision energy 13~TeV, rapidity $y = 0$ and $\hat r = r_p$.  It is
      compared to the dash-dotted line showing the survival factor from
      Ref.~\cite{1410.2983}.\protect\footnotemark{}
      The lines were extracted from
      the papers with the help of Ref.~\cite{web-plot-digitizer}.
    }
    \label{f:comparison}
  \end{minipage}
\end{figure}

The luminosity of photons appearing in ultraperipheral collisions of protons
with the energy 13~TeV is plotted in Fig.~\ref{f:survival} along with the
survival factor. There is little point in going beyond the presented range of
invariant masses, because a proton will not survive emitting a photon with the
energy above $\sim 1.4$~TeV~\cite{1806.07238}. Although the photons in UPC
are polarized, the difference in the luminosities for different polarization
arrangements is at the level of a few percent, resulting in negligible
difference in the cross section for the production of a pair of charged
fermions, see Fig.~\ref{f:chargino}.

Taking into account the survival factor improves the agreement between the
theoretical description of the ATLAS experiment and the experimental data, see
Fig.~\ref{f:atlas}.

In Fig.~\ref{f:comparison}, we compare our calculations with the results
obtained in Refs.~\cite{1508.02718, 1410.2983}. The reduction of the cross
section according to Ref.~\cite{1410.2983} is much larger due to the restriction
$b_i > r_p$ (or $\hat r = r_p$) imposed in Ref.~\cite{1410.2983}. We did not
impose such a restriction since diminishing of the photon flux at $b_i < r_p$ is
taken into account by the elastic form factor, and the produced particles do not
interact strongly with the  protons. The numerical difference with the results
obtained in~\cite{1508.02718} is marginal.

We are grateful to V.~A.~Khoze for the comments on photons polarizations in UPC
collisions and the references to previous work. The authors were supported by
the Russian Science Foundation grant No 19-12-00123.

\footnotetext{
  Note that the constraint $\hat r = r_p$ applied in Ref.~\cite{1410.2983} is
  incorrect in the case of muon pair production, see the discussion after
  Eq.~\eqref{pp->ppX-ip} and Ref.~\cite{2104.13392}.
}

\clearpage

\newcommand{\arxiv}[1]{\href{http://arxiv.org/abs/#1}{arXiv:\nolinebreak[3]#1}}


\begin{thebibliography}{99}
\bibitem{prep163-299}
C.~A.~Bertulani, G.~Baur.
Electromagnetic Processes in relativistic heavy ion collisions.
Phys.Rept.~163, 299 (1988).

\bibitem{hep-ph-0112211}
G.~Baur, K.~Hencken, D.~Trautmann, S.~Sadovsky, Yu.~Kharlov.
Coherent $\gamma \gamma$ and $\gamma A$ interactions in very peripheral collisions at relativistic ion colliders.
Phys.Rept.~364, 359 (2002).
\href{http://arxiv.org/abs/hep-ph/0112211}{arXiv:\nolinebreak[3]hep-ph/0112211}

\bibitem{hep-ph-0112239}
G.~Baur.
Physics opportunities in ultraperipheral heavy ion collisions at LHC.
~(2001).
Workshop on electromagnetic probes of fundamental physics, p. 183 (2001).
\href{http://arxiv.org/abs/hep-ph/0112239}{arXiv:\nolinebreak[3]hep-ph/0112239}

\bibitem{hep-ex-0201034}
G.~Baur, C.A.~Bertulani, M.~Chiu, I.F.~Ginzburg, {\it et.al}.
Hot topics in ultra-peripheral ion collisions.
Workshop on electromagnetic probes of fundamental physics, p. 235 (2002).
\href{http://arxiv.org/abs/hep-ex/0201034}{arXiv:\nolinebreak[3]hep-ex/0201034}

\bibitem{hep-ph-0304301}
L.~Frankfurt, M.~Strikman, M.~Zhalov.
Coherent photoproduction from nuclei.
Acta Phys.Polon.~B34, 3215 (2003).
\href{http://arxiv.org/abs/hep-ph/0304301}{arXiv:\nolinebreak[3]hep-ph/0304301}

\bibitem{nucl-ex-0502005}
C.~A.~Bertulani, S.~R.~Klein, J.~Nystrand.
Physics of ultra-peripheral nuclear collisions.
Ann.Rev.Nucl.Part.Sci.~55, 271 (2005).
\href{http://arxiv.org/abs/nucl-ex/0502005}{arXiv:\nolinebreak[3]nucl-ex/0502005}

\bibitem{hep-ph-0611042}
J.~Nystrand.
Ultra-peripheral collisions of heavy ions at RHIC and the LHC.
Nucl.Phys.~A787, 29 (2007).
\href{http://arxiv.org/abs/hep-ph/0611042}{arXiv:\nolinebreak[3]hep-ph/0611042}

\bibitem{0706.3356}
A.~J.~Baltz, G.~Baur, D.~d'Enterria, L.~Frankfurt {\it et.~al.}
The physics of ultraperipheral collisions at the LHC.
Phys.Rept.~458, 1 (2008).
\arxiv{0706.3356}

\bibitem{0810.1400}
G.~Baur.
Coherent photon-photon interactions in very peripheral relativistic heavy ion
collisions.
Eur.Phys.J.~D55, 265 (2009).
\arxiv{0810.1400}

\bibitem{1104.0571}
M.~K\l{}usek-Gawenda, A.~Szczurek.
Exclusive production of large invariant mass pion pairs in ultraperipheral
ultrarelativistic heavy ion collisions.
Phys.Lett.~B700, 322 (2011).
\arxiv{1104.0571}

\bibitem{1404.0896}
A.~Szczurek.
Peripheral, ultrarelativistic production of particles in heavy ion collisions.
Acta~Phys.Polon.~B45, 1597 (2014).
\arxiv{1404.0896}

\bibitem{1601.07001}
M.~K\l{}usek-Gawenda, P.~Lebiedowicz, A.~Szczurek.
Light-by-light scattering in ultraperipheral PbPb collisions at energies available at the CERN Large Hadron Collider.
Phys.Rev.~C93, 044907 (2016).
\arxiv{1601.07001}

\bibitem{1607.05095}
M.~K\l{}usek-Gawenda, A.~Szczurek.
Double scattering production of two positron-electron pairs in ultraperipheral
heavy-ion collisions.
Phys.Lett.~B763, 416 (2016).
\arxiv{1607.05095}

\bibitem{1610.06647}
M.~B.~Gay Ducati, F.~Kopp, M.~V.~T.~Machado, S.~Martins.
Photoproduction of Upsilon states in ultraperipheral collisions at the CERN
Large Hadron Collider with the color dipole approach.
Phys.Rev.~D94, 094023 (2016).
\arxiv{1610.06647}

\bibitem{1708.09836}
M.~K\l{}usek-Gawenda, P.~Lebiedowicz, O.~Nachtmann, A.~Szczurek.
From the $\gamma \gamma \to p \bar p$ reaction to the production of $p \bar p$
pairs in ultraperipheral ultrarelativistic heavy-ion collisions at the LHC.
Phys.Rev.~D96, 094029 (2017).
\arxiv{1708.09836}

\bibitem{prd42-3690}
R.~N.~Cahn, J.~D.~Jackson.
Realistic equivalent-photon yields in heavy-ion collisions.
Phys.~Rev.~D42, 3690 (1990).

\bibitem{prc47-2308}
M.~Vidovi\'c, M.~Greiner, C.~Best, G.~Soff.
Impact-parameter dependence of the electromagnetic particle production in
ultrarelativistic heavy-ion collisions.
Phys.Rev.~C47, 2308 (1993).

\bibitem{prc48-2011}
M.~Vidovi\'c, M.~Greiner, G.~Soff.
Electromagnetic dissociation of Pb nuclei in peripheral ultrarelativistic
heavy-ion collisions.
Phys.Rev.~C48, 2011 (1993).

\bibitem{1806.07238}
M.~Vysotsky, E.~Zhemchugov.
Equivalent photons in proton-proton and ion-ion collisions at the Large Hadron Collider.
Physics-Uspekhi 62, 910 (2019).
\arxiv{1806.07238}

\bibitem{pr119-1105}
F.~J.~Ernst, R.~G.~Sachs, K.~C.~Wali.
Electromagnetic form factors of the nucleon.
Phys.~Rev.~119, 1105 (1960).

\bibitem{prep550-1}
S.~Pacetti, R.~B.~Ferroli, E.~Tomasi-Gustafsson.
Proton electromagnetic form factors: basic notions, present achievements and
future perspectives.
Phys.Rep.~550, 1 (2015).
\bibitem{1507.07956}
P.~J.~Mohr, D.~B.~Newell, B.~N.~Taylor.
CODATA recommended values of the fundamental physical constants: 2014.
Rev.~Mod.~Phys.~88, 035009 (2016).
\arxiv{1507.07956}

\bibitem{1307.6227}
The A1 collaboration.
Electric and magnetic form factors of the proton.
Phys.Rev.~C90, 015206 (2014).
\arxiv{1307.6227}

\bibitem{npa518-786}
G.~Baur, L.~G.~Ferreira~Filho.
Coherent particle production at relativistic heavy-ion colliders including
strong absorption effects.
Nucl.~Phys.~A518, 786 (1990).

\bibitem{hep-ph-0608271}
L.~Frankfurt, C.~E.~Hyde-Wright, M.~Strikman, C.~Weiss.
Generalized parton distributions and rapidity gap survival in exclusive diffractive $p p$ scattering.
Phys.Rev.~D75, 054009 (2007).
\href{http://arxiv.org/abs/hep-ph/0608271}{arXiv:\nolinebreak[3]hep-ph/0608271}

\bibitem{1112.3243}
V.~A.~Schegelsky, M.~G.~Ryskin.
Diffraction cone shrinkage speed up with the collision energy.
Phys.Rev.~D85, 094024 (2012).
\arxiv{1112.3243}

\bibitem{1408.5778}
The ATLAS Collaboration.
Measurement of the total cross section from elastic scattering in $pp$
collisions at $\sqrt{s} = 7$~TeV with the ATLAS detector.
Nucl.Phys.~B889, 486 (2014).
\arxiv{1408.5778}

\bibitem{1607.06605}
The ATLAS Collaboration.
Measurement of the total cross section from elastic scattering in $pp$
collisions at $\sqrt{s} = 8$~TeV with the ATLAS detector.
Phys.Lett.~B761, 158 (2016).
\arxiv{1607.06605}

\bibitem{hep-ph-0010163}
V.~A.~Khoze, A.~D.~Martin, R.~Orava, M.~G.~Ryskin.
Luminosity measuring processes at the LHC.
Eur.~Phys.~J.~C19, 313 (2001).
\arxiv{hep-ph/0010163}

\bibitem{hep-ph-0201301}
V.~A.~Khoze, A.~D.~Martin, M.~G.~Ryskin.
Photon-exchange processes at hadron colliders as a probe of the dynamics of
diffraction.
Eur.~Phys.~J.~C24, 459 (2002).
\arxiv{hep-ph/0201301}

\bibitem{1405.0018}
L.~A.~Harland-Lang. V.~A.~Khoze, M.~G.~Ryskin, W.~J.~Stirling.
Central exclusive production within the Durham model: a review.
Int.~J.~Mod.~Phys.~A29, 1430031 (2014).
\arxiv{1405.0018}

\bibitem{1508.02718}
L.~A.~Harland-Lang, V.~A.~Khoze, M.~G.~Ryskin.
Exclusive physics at the~LHC with SuperChic~2.
Eur.Phys.J.~C76, 9 (2016).
\arxiv{1508.02718}

\bibitem{1710.11505}
V.~A.~Khoze, A.~D.~Martin, M.~G.~Ryskin.
Multiple interactions and rapidity gap survival.
J.~Phys.~G45, 053002 (2018).
\arxiv{1710.11505}

\bibitem{2007.12704}
L.~A.~Harland-Lang, M.~Tasevsky, V.~A.~Khoze, M.~G.~Ryskin.
A new approach to modelling elastic and inelastic photon-initiated production at
the LHC: SuperChic~4.
Eur.~Phys.~J.~C80, 925 (2020).
\arxiv{2007.12704}

\bibitem{pr46.1087}
G.~Breit, J.~A.~Wheeler.
Collision of two light quanta.
Phys.~Rev.~46, 1087 (1934).

\bibitem{1906.08568}
S.~I.~Godunov, V.~A.~Novikov, A.~N.~Rozanov, M.~I.~Vysotsky, E.~V.~Zhemchugov.
Quasistable charginos in ultraperipheral proton-proton collisions at the LHC.
JHEP~2001, 143 (2020).
\arxiv{1906.08568}

\bibitem{atlas-tdr-024-2015}
The ATLAS Collaboration.
ATLAS Forward Proton phase I upgrade. Technical Design Report.
CERN-LHCC-2015-009, ATLAS-TDR-024-2015.

\bibitem{totem-tdr-003}
The CMS and TOTEM Collaborations.
CMS-TOTEM Precision Proton Spectrometer. Technical Design Report.
CERN-LHCC-2014-021, TOTEM-TDR-003.

\bibitem{1708.04053}
The ATLAS Collaboration.
Measurement of the exclusive $\gamma \gamma \to \mu^+ \mu^-$ process in proton-proton collisions at $\sqrt{s} = 13$~TeV with the ATLAS detector.
Phys.Lett. B~777, 303 (2018).
\arxiv{1708.04053}

\bibitem{0803.0883}
M.~B\"ahr, S.~Gieseke, M.A.~Gigg, D.~Grellscheid, {\it et.al}.
Herwig++ physics and manual.
Eur.Phys.J.~C58, 639 (2008).
\href{http://arxiv.org/abs/0803.0883}{arXiv:\nolinebreak[3]0803.0883}

\bibitem{1512.01178}
J.~Bellm, S.~Gieseke, D.~Grellscheid, S.~Pl\"atzer, {\it et.al}.
Herwig 7.0/Herwig++ 3.0 release note.
Eur.Phys.J.~C76, 196 (2016).
\href{http://arxiv.org/abs/1512.01178}{arXiv:\nolinebreak[3]1512.01178}

\bibitem{1410.2983}
M.~Dyndal, L.~Schoeffel.
The role of finite-size effects on the spectrum of equivalent photons in
proton-proton collisions at the LHC.
Phys.Lett.B~741, 66 (2015).
\arxiv{1410.2983}

\bibitem{2104.13392}
L.~A.~Harland-Lang, V.~A.~Khoze, M.~G.~Ryskin.
Elastic photon-initiated production at the LHC: the role of hadron-hadron
interactions.
\arxiv{2104.13392}

\bibitem{web-plot-digitizer}
A.~Rohatgi.
WebPlotDigitizer---Web Based Plot Digitizer. \raggedright
\href{https://automeris.io/WebPlotDigitizer}{https://automeris.io/WebPlotDigitizer}


\end{thebibliography}
\end{document}